\documentclass[amsmath,amssymb,aps,prd,reprint,twocolumn,showkeys,showpacs]{revtex4-1}

\usepackage{xcolor}
\usepackage{amsmath}
\usepackage{graphicx}
\usepackage{dcolumn}
\usepackage{bm}
\usepackage{epstopdf}
\printfigures
\usepackage{epsfig}

\usepackage{subfigure}
\usepackage{subfig}

\usepackage{booktabs}

\begin{document}

\title{Dynamical aspects for scalar fields coupled to cubic contractions of the Riemann tensor}

\author{Mihai Marciu}
\email{mihai.marciu@drd.unibuc.ro}
\affiliation{Faculty of Physics, University of Bucharest, 405 Atomi\c{s}tilor, POB MG-11, RO-077125, Bucharest-M\u{a}gurele, Romania}

\begin{abstract}
The paper studies a new type of dark energy, a scalar field with positive or negative kinetic energy, generically coupled to a term which is composed by specific contractions of the Riemann tensor. After presenting the resulting field equations, we have analyzed the physical characteristics of the corresponding model by implementing the linear stability theory. In the case of an exponential coupling function and exponential potential energy we have deduced the phase space characteristics, analyzing the critical points obtained which can represent specific eras in the evolution of the Universe. The analytical study is showing that this model can represent a feasible cosmological setup, having various epochs which correspond to stiff--fluid, matter domination, and dark energy eras, pointing towards the emergence of the accelerated expansion as a geometrical effect. \end{abstract}

\maketitle

\newpage 

\section{Introduction}
\par
The dynamics of the Universe at the large scale organization have always been a captivating area of study for physicists in modern times. The evolution at the large scale structure is mainly described by the general relativity theory, a theoretical framework capable of characterizing the gravitational interaction. Although the general relativity represents a solid element in the foundation of modern physics, it suffers from a series of inconsistencies and limitations concerning the juxtaposition with observations acquired by different astrophysical studies \cite{Bull:2015stt,Joyce:2014kja, Weinberg:1988cp, Baker:2014zba}. One of the key issues is related to the existence of the dark energy problem, a new phenomenon discovered almost twenty years ago which still possesses an enigmatic question \cite{Amendola:2012ky, Calder:2007ci, Ade:2015rim,Huterer:2002hy,Kunz:2012aw,Motta:2013cwa,Baldi:2012ky,Josset:2016vrq}. This phenomenon is related to the accelerated expansion of the Universe and represents an open problem in the modern cosmology \cite{Peebles:2002gy}. The most simple theoretical setup which can explain the accelerated expansion of the Universe is described by the $\Lambda$CDM model \cite{spergel}, a theory which includes the addition of a cosmological constant to the Einstein--Hilbert component. In this case, the gravitational theory has a reduced complexity and cannot explain various evolutionary aspects \cite{Astashenok:2012qn, Popolo:2015lyb,Fielder:2018szt} related to the behavior of the Universe at large scale, describing inconsistently the dark sector. It can be considered only as an effective theory acting more as an approximate framework which needs to be further revised in order to explain the astrophysical observations at large and small scales. In the recent years many observational studies have shown the inconsistencies and limitations of the $\Lambda$CDM model \cite{DelPopolo:2016emo,Martin:2012bt, Copi:2010na}.
\par 
Another key issue for the $\Lambda$CDM model is related to the cosmic coincidence problem, a specific issue which characterizes the late time evolution of the Universe \cite{Martin:2012bt, Franca:2005dh, Rivera:2016zzr, Sivanandam:2012ty, Velten:2014nra,Jamil:2008nta}. From a theoretical perspective these issues presents serious inconsistencies of the $\Lambda$CDM cosmological model which needs to be addressed in order to build a more fundamental theory for the gravitational interaction at large scales. Moreover, various studies have added new problems and limitations for the $\Lambda$CDM model at small scales \cite{Popolo:2015lyb,DelPopolo:2016emo}, opening new viable theoretical directions. In scalar tensor theories, a possible extension of the Einstein--Hilbert action includes the addition of one or more scalar fields, minimally or non--minimally coupled to various geometrical invariants \cite{Bahamonde:2017ize}. The viability of the scalar fields in the form of quintessence \cite{Sami:2013ssa, Novosyadlyj_2013} or phantom dark energy models \cite{Ludwick:2015dba} have been analyzed, addressing many theoretical and observational directions in scalar tensor theories of gravitation \cite{Copeland:2006wr,Tian:2019enx,Avelino:2004vy}.

\par 
In the scalar tensor cosmological models of gravity the Einsteinian cubic gravity can be regarded as a notable extension of the general relativity proposed recently by Bueno and Cano \cite{Bueno:2016xff}. This theory involves the addition of a component which includes particular contractions of the Riemann tensor in the cubic order. The physical properties of the Einsteinian cubic gravity have been studied in the recent years in various directions of study \cite{Bueno:2018yzo,Bueno:2018xqc, Jiang:2019kks, Bueno:2020odt, Pookkillath:2020iqq}. The extension of the Einsteinian cubic gravity towards a generic theory which depends of the specific cubic term have appeared in Ref.~\cite{Erices:2019mkd}, showing the late time evolution of the equation of state closer to the observational interval. The cosmological solutions based on different black hole types in the case of Einsteinian cubic gravity and particular generalizations have been analyzed in  \cite{Hennigar:2016gkm, Bueno:2016lrh, Bueno:2017sui,Feng:2017tev,Emond:2019crr, Cano:2019ozf, Burger:2019wkq, Frassino:2020zuv, KordZangeneh:2020qeg}. On the other hand, the wormhole solutions for the cubic gravity case have been analyzed from a theoretical point of view \cite{Mehdizadeh:2019qvc}. Using the same considerations, the relation of the cubic gravity with the inflationary dynamics have been addressed in Refs.~\cite{Arciniega:2019oxa, Arciniega:2018tnn, Arciniega:2018fxj}. Then, from a dynamical perspective the viability of the generalized cubic gravity have been studied also by considering the linear stability theory for exponential and power law types \cite{Marciu:2020ysf}. The case where a cosmological constant is added to the Einsteinian cubic gravity has been investigated using the dynamical system analysis, revealing the phase space properties and various fundamental aspects \cite{Quiros:2020uhr}.

\par 
In this paper we shall investigate the case of scalar fields with both positive and negative kinetic terms endowed with potential energy, non--minimally coupled in a generic manner with a term which contains specific contractions of the Riemann tensor in the third order \cite{Erices:2019mkd,Bueno:2016xff}. The physical consequences of the non--minimal coupling with specific cubic contractions of the Riemann tensor will be discussed by adopting the linear stability theory \cite{Bahamonde:2017ize} in the case where both the coupling function and the potential energy have an exponential behavior. The main aim of the investigation is related to the analysis of the phase space structure and the theoretical viability of such a cosmological model in an attempt of explaining the late time evolution of the Universe at the level of background dynamics, offering a possible solution to the existence of the matter epoch and the accelerated expansion era near the de--Sitter regime without fine--tuning.

\par 
The plan of the paper is the following: in Sec.~\ref{sec:adoua} we present the action corresponding to our model and the resulting field equations, the Klein--Gordon and the modified Friedmann relations. Then, in Sec.~\ref{sec:atreia} we analyze the physical characteristics of our model by using the linear stability method, in the case of an exponential coupling function and potential energy, discussing the viability of the proposed cosmological model. Lastly, we present the summary of our study and the final concluding remarks in Sec.~\ref{sec:concluzii}.

\section{The model}
\label{sec:adoua} 

In this section, we shall discuss the action and the resulting field equations for the cosmological model. Before proceeding to the presentation of the action for the cosmological model we specify the convention of the Robertson--Walker metric associated to the FRW background:
\begin{equation}
\label{metrica}
ds^2=-dt^2+a^2(t) \delta_{ju}dx^j dx^u.
\end{equation}
\par 
In this background we denote with $a(t)$ the time dependent cosmic scale factor which describes a homogeneous and isotropic Universe. We shall extend the Einstein--Hilbert action by adding a scalar field which can be canonical or non--canonical, having a positive or negative kinetic energy, with the associated sign embedded into the value of the $\epsilon$ parameter. In the present cosmological model the scalar field is endowed with a potential energy term, having a non--minimal coupling with a term which contains specific contractions of the Riemann tensor up to the cubic order. The resulting action which can describe such a cosmological scenario has the following form \cite{Erices:2019mkd}:
\begin{equation}
\label{actiune}
S=S_m+\int d^4x \sqrt{-g} \Bigg( \frac{R}{2}-\frac{\epsilon}{2} g^{\mu\nu}\partial_{\mu}\phi\partial_{\nu}\phi-V(\phi)+f(\phi)P\Bigg),
\end{equation}
where  \cite{Bueno:2016xff} 
\begin{multline}
P=\beta_1 R_{\mu\quad\nu}^{\quad\rho\quad\sigma}R_{ \rho\quad\sigma}^{\quad \gamma\quad\delta}R_{\gamma\quad\delta}^{\quad\mu\quad\nu}+\beta_2 R_{\mu\nu}^{\rho\sigma}R_{\rho\sigma}^{\gamma\delta}R_{\gamma\delta}^{\mu\nu}
\\+\beta_3 R^{\sigma\gamma}R_{\mu\nu\rho\sigma}R_{\quad\quad\gamma}^{\mu\nu\rho}+\beta_4 R R_{\mu\nu\rho\sigma}R^{\mu\nu\rho\sigma}+\beta_5 R_{\mu\nu\rho\sigma}R^{\mu\rho}R^{\nu\sigma}
\\+\beta_6 R_{\mu}^{\nu}R_{\nu}^{\rho}R_{\rho}^{\mu}+\beta_7 R_{\mu\nu}R^{\mu\nu}R+\beta_8 R^3.
\end{multline}
Next, considering the following inter--relations between various parameters associated to the cubic term \cite{Erices:2019mkd, Bueno:2016xff}:
\begin{equation}
\beta_7=\frac{1}{12}\big[3\beta_1-24\beta_2-16\beta_3-48\beta_4-5\beta_5-9\beta_6\big],
\end{equation}
\begin{equation}
\beta_8=\frac{1}{72}\big[-6\beta_1+36\beta_2+22\beta_3+64\beta_4+5\beta_5+9\beta_6\big],
\end{equation}
\begin{equation}
\beta_6=4\beta_2+2\beta_3+8\beta_4+\beta_5,
\end{equation}
\begin{equation}
\bar{\beta}=(-\beta_1+4\beta_2+2\beta_3+8\beta_4).
\end{equation}
we can show that for the current cosmological background the cubic component represents a second order term, with the following expression:
\begin{equation}
\label{PP}
P=6\bar{\beta}H^4 (2H^2+3\dot{H}).
\end{equation}
\par
For the action proposed in the relation \eqref{actiune} the dynamics of the corresponding gravitational sector \cite{Erices:2019mkd} are described by the following modified Friedmann relations: 
\begin{equation}
3H^2=\rho_m+\rho_{\phi},
\end{equation}
\begin{equation}
3H^2+2\dot{H}=-p_m-p_{\phi},
\end{equation}
where
\par 
\begin{equation}
\rho_{\phi}=\frac{1}{2}\epsilon\dot{\phi}^2+V(\phi)+6 \beta f(\phi) H^6-18 \beta H^5 \frac{df(\phi)}{d\phi}\dot{\phi},
\end{equation}
\begin{multline}
p_{\phi}=\frac{1}{2}\epsilon\dot{\phi}^2-V(\phi)-6 \beta f(\phi) H^6-12 \beta f(\phi) H^4 \dot{H}
\\+12 \beta H^5 \frac{df(\phi)}{d\phi}\dot{\phi}+24 \beta H^3 \frac{df(\phi)}{d\phi}\dot{H}\dot{\phi}
\\
+6 \beta H^4 \dot{\phi}^2\frac{d^2f(\phi)}{d\phi^2}+6 \beta H^4 \frac{df(\phi)}{d\phi}\ddot{\phi}.
\end{multline}
We can further define the equation of state for the scalar field
\begin{equation}
w_{\bf{\phi}}=\frac{p_{\phi}}{\rho_{\phi}},
\end{equation}
and the effective equation of state for the scalar field model  non--minimally coupled to the specific cubic contractions of the Riemann tensor:
\begin{equation}
w_{\bf{eff}}=\frac{p_m+p_{\phi}}{\rho_{m}+\rho_{\phi}}=-1-\frac{2}{3}\frac{\dot{H}}{H^2}.
\end{equation}
\par 
In this case the Klein--Gordon relation can be written as:
\begin{equation}
\epsilon(\ddot{\phi}+3 H \dot{\phi})+\frac{dV(\phi)}{d\phi}-6 \beta H^4 (2 H^2+3 \dot{H})\frac{df(\phi)}{d\phi}=0.
\end{equation}
\par 
Due to the fact that the action doesn't contains any terms where the geometry is coupled with the energy momentum tensor we can expect that the dark energy component is characterized by a standard continuity equation of the following type 
\begin{equation}
\dot{\rho_{\phi}}+3H(\rho_{\phi}+p_{\phi})=0.
\end{equation}
\par 
For the matter component ($p_m=w_m\rho_m$) described by the $S_m$ term in the total action \eqref{actiune} we also have a continuity equation :
\begin{equation}
\dot{\rho_{m}}+3H(\rho_{m}+p_{m})=0.
\end{equation}
\par 
Lastly, we can define in the usual manner the density parameters for the two constituents, the matter component
\begin{equation}
\Omega_m=\frac{\rho_{m}}{3H^2},
\end{equation}
and the dark energy sector as the scalar field non--minimally coupled to cubic contractions of the Riemann tensor
\begin{equation}
\Omega_{\phi}=\frac{\rho_{\phi}}{3H^2},
\end{equation}
obtaining the ordinary constraint
\begin{equation}
\Omega_m+\Omega_{\phi}=1.
\end{equation}

\section{The phase space and the corresponding physical aspects}
\label{sec:atreia} 
\par 

In this section, we start with the transformation of the equations corresponding to the cosmological model from the cosmic time $t$ to $N=log(a)$, linearizing the dynamical equations. In what follows we introduce the specific variables:
\begin{equation}
    x=\frac{\dot{\phi}}{\sqrt{3}H},
\end{equation}
\begin{equation}
    y=\frac{\sqrt{V(\phi)}}{\sqrt{3}H},
\end{equation}
\begin{equation}
    z=2 \beta f(\phi) H^4,
\end{equation}
\begin{equation}
    s=\Omega_m=\frac{\rho_m}{3 H^2},
\end{equation}
rewriting the first Friedmann equation into the following relation:
\begin{equation}
\label{eqfr1111}
    1=s+\frac{1}{2}\epsilon x^2+y^2+z-3 \sqrt{3} \alpha z x.
\end{equation}
For the Klein--Gordon equation, we can write:
\begin{equation}
   \epsilon \ddot{\phi}+\epsilon 3 \sqrt{3} H^2 x-\lambda 3 H^2 y^2-6 H^2 \alpha z-9 \dot{H} \alpha z=0.
\end{equation}
Note that in this case we have considered an exponential coupling and potential energy, described by 
\begin{equation}
   V(\phi)=V_0 e^{-\lambda \phi},
\end{equation}
\begin{equation}
   f(\phi)=f_0 e^{\alpha \phi},
\end{equation}
with $V_0,f_0,\alpha,\lambda$ constant parameters. The second Friedmann relation which characterizes the dynamical acceleration of the model has the following formula:
\begin{multline}
   -3 H^2-2 \dot{H}=s w_m 3 H^2+\frac{1}{2}\epsilon x^2 H^2 3
   -3 H^2 y^2-3 H^2 z
   \\-6 z \dot{H} +\frac{12}{2}\alpha z x H^2 \sqrt{3}+12 \alpha z x \sqrt{3} \dot{H}
   +9 x^2 H^2 \alpha^2 z
   \\+3 \alpha z \ddot{\phi}.
\end{multline}
\par 
Then, we can write the explicit dynamical system where we denote with $'$ the differentiation withe respect to $N$, considering $N=log(a)$:
\begin{equation}
  x'=\frac{dx}{dN}=\frac{1}{\sqrt{3}}\frac{\ddot{\phi}}{H^2}-x \frac{\dot{H}}{H^2},
\end{equation}
\begin{equation}
  y'=\frac{dy}{dN}=-\frac{\sqrt{3}}{2} \lambda y x - y \frac{\dot{H}}{H^2},
\end{equation}
\begin{equation}
  z'=\frac{dz}{dN}=\alpha z x \sqrt{3}+4 z \frac{\dot{H}}{H^2}.
\end{equation}
At this point, we can note that the dynamical system associated to the present cosmological model is completely autonomous and can be characterized by the specific method which uses the linear stability theory. Notice also that in our case the system becomes a three dimensional one by using the Friedmann constraint expressed in eq.~\ref{eqfr1111}. With all of these considerations, our dynamical system reduces in the final form to:
\onecolumngrid
\begin{multline}
    x'=\frac{1}{12 z \epsilon  \left(2 \sqrt{3} \alpha  x-1\right)+54 \alpha ^2 z^2+4 \epsilon } \cdot \Big( -3 x^3 \epsilon ^2 w_m+27 \sqrt{3} \alpha  x^2 z \epsilon  w_m-6 x y^2 \epsilon  w_m-162 \alpha ^2 x z^2 w_m-6 x z \epsilon  w_m+6 x \epsilon  w_m
    \\+18 \sqrt{3} \alpha  y^2 z w_m+18 \sqrt{3} \alpha  z^2 w_m-18 \sqrt{3} \alpha  z w_m+18 \alpha ^2 x^3 z \epsilon +3 x^3 \epsilon ^2-54 \sqrt{3} \alpha ^3 x^2 z^2-87 \sqrt{3} \alpha  x^2 z \epsilon +90 \alpha  \lambda  x y^2 z-6 x y^2 \epsilon +72 \alpha ^2 x z^2
    \\+30 x z \epsilon -6 x \epsilon +4 \sqrt{3} \lambda  y^2+18 \sqrt{3} \alpha  y^2 z-12 \sqrt{3} \lambda  y^2 z-6 \sqrt{3} \alpha  z^2-10 \sqrt{3} \alpha  z \Big),
\end{multline}

\begin{multline}
    y'=\frac{1}{12 z \epsilon  \left(2 \sqrt{3} \alpha  x-1\right)+54 \alpha ^2 z^2+4 \epsilon} \cdot \Big( -3 x^2 y \epsilon ^2 w_m+18 \sqrt{3} \alpha  x y z \epsilon  w_m-6 y^3 \epsilon  w_m-6 y z \epsilon  w_m+6 y \epsilon  w_m
    +18 \alpha ^2 x^2 y z \epsilon 
    \\-36 \alpha  \lambda  x^2 y z \epsilon +3 x^2 y \epsilon ^2-27 \sqrt{3} \alpha ^2 \lambda  x y z^2-6 \sqrt{3} \alpha  x y z \epsilon +6 \sqrt{3} \lambda  x y z \epsilon -2 \sqrt{3} \lambda  x y \epsilon +18 \alpha  \lambda  y^3 z-6 y^3 \epsilon +36 \alpha ^2 y z^2-6 y z \epsilon +6 y \epsilon \Big),
\end{multline}

\begin{multline}
    z'=\frac{1}{6 z \epsilon  \left(2 \sqrt{3} \alpha  x-1\right)+27 \alpha ^2 z^2+2 \epsilon} \cdot \Big( 6 x^2 z \epsilon ^2 w_m-36 \sqrt{3} \alpha  x z^2 \epsilon  w_m+12 y^2 z \epsilon  w_m+12 z^2 \epsilon  w_m-12 z \epsilon  w_m-6 x^2 z \epsilon ^2
    \\+27 \sqrt{3} \alpha ^3 x z^3+6 \sqrt{3} \alpha  x z^2 \epsilon +2 \sqrt{3} \alpha  x z \epsilon -36 \alpha  \lambda  y^2 z^2+12 y^2 z \epsilon -72 \alpha ^2 z^3+12 z^2 \epsilon -12 z \epsilon \Big) .
\end{multline}
Next, we also add the relation for the effective equation of state,
\begin{multline}
    w_{\bf{eff}}=\frac{x^2 \epsilon  \left(-\epsilon  w_m+6 \alpha ^2 z+\epsilon \right)+2 \sqrt{3} \alpha  x z \epsilon  \left(3 w_m-7\right)-2 y^2 \left(\epsilon  w_m-3 \alpha  \lambda  z+\epsilon \right)-2 z \epsilon  w_m+2 \epsilon  w_m-15 \alpha ^2 z^2+4 z \epsilon }{6 z \epsilon  \left(2 \sqrt{3} \alpha  x-1\right)+27 \alpha ^2 z^2+2 \epsilon }.
\end{multline}

\begin{table}[h]
\begin{center}
\begin{tabular}{ c |c| c |c |c| c | c}
 Cr.Point & x & y & z & $\Omega_m$ & $w_{\bf{eff}}$ & Eigenvalues \\ 
 $P_1$ & 0 & 0 & 0 & 1 & $w_m$ & $\Big[\frac{3}{2} \left(w_m-1\right),-6 \left(w_m+1\right),\frac{3}{2} \left(w_m+1\right)\Big]$ \\  
 $P_2$ & 0 & $\frac{\sqrt{2} \sqrt{\alpha }}{\sqrt{2 \alpha -\lambda }}$ & $\frac{\lambda }{\lambda -2 \alpha }$ & 0 & -1 & See discussion. \\ 
 $P_3$ & $\frac{2 \sqrt{3} \left(w_m+1\right)}{\alpha }$ & 0 & $\frac{6 \epsilon  \left(w_m^2-1\right)}{\alpha ^2 \left(9 w_m+5\right)}$ & $\frac{6 \epsilon  \left(9 w_m^3-6 w_m^2-37 w_m-22\right)}{\alpha ^2 \left(9 w_m+5\right)}+1$ & $w_m$ & See discussion.\\
 $P_4$ & $-\frac{\sqrt{2}}{\sqrt{\epsilon }}$ & 0 & 0 & 0 & 1 & $\Big[-\frac{\sqrt{6} \alpha }{\sqrt{\epsilon }}-12,3-3 w_m,\frac{\sqrt{\frac{3}{2}} \lambda }{\sqrt{\epsilon }}+3\Big]$\\
 $P_5$ & $\frac{\sqrt{2}}{\sqrt{\epsilon }}$ & 0 & 0 & 0 & 1 & $\Big[ \frac{\sqrt{6} \alpha }{\sqrt{\epsilon }}-12,3-3 w_m,3-\frac{\sqrt{\frac{3}{2}} \lambda }{\sqrt{\epsilon }} \Big]$\\
 $P_6$ & $\frac{\sqrt{3} \left(w_m+1\right)}{\lambda }$ & $-\frac{\sqrt{\frac{3}{2}} \sqrt{\epsilon -\epsilon  w_m^2}}{\lambda }$ & $0$ & $1-\frac{3 (w_m+1) \epsilon }{\lambda ^2}$ & $w_m$ & See discussion. \\
 $P_7$ & $\frac{\sqrt{3} \left(w_m+1\right)}{\lambda }$ & $\frac{\sqrt{\frac{3}{2}} \sqrt{\epsilon -\epsilon  w_m^2}}{\lambda }$ & 0 & $1-\frac{3 \epsilon  \left(w_m+1\right)}{\lambda ^2}$ & $w_m$ & See discussion.\\
 $P_8$ & $\frac{\lambda }{\sqrt{3} \epsilon }$ & $\sqrt{1-\frac{\lambda ^2}{6 \epsilon }}$ & 0 & 0 & $\frac{\lambda ^2}{3 \epsilon }-1$ & $\Big[ \frac{\alpha  \lambda }{\epsilon }-\frac{2 \lambda ^2}{\epsilon },\frac{\lambda ^2}{2 \epsilon }-3,-3 w_m+\frac{\lambda ^2}{\epsilon }-3 \Big]$\\
\end{tabular}
\end{center}
\caption{\label{tab:table1} The critical points in the phase space and some of the physical characteristics.}
\end{table}
\twocolumngrid
In the Table~\ref{tab:table1} we show the critical points attached to the present cosmological scenario which includes the addition to the Einstein Hilbert action of a scalar field (with positive or negative kinetic energy), non--minimally coupled to a specific invariant which is embedding various cubic contractions of the Riemann tensor. Notice that the exact relations for the eigenvalues in the case of complex cosmological solutions are presented as a discussion in the text. In the next paragraphs we shall describe each critical point in detail, analyzing the possible viability of the corresponding cosmological solutions from a physical and dynamical point of view.  
\par 
The first critical point $P_1$ represents the origin of the phase space, a matter dominated epoch ($\Omega_m=1$), without any influence from any kinetic/potential features of the scalar field, or non--minimal coupling effects. At this point, the effective equation of state describes a matter content ($w_{\bf{eff}}=w_m$). If we take into account a dust scenario where the matter component act as a cold gas without pressure ($w_m=0$), then the solution is always saddle from a dynamical point of view.
\par
\begin{figure*}[t]
\centering
\includegraphics[width=0.4\textwidth]{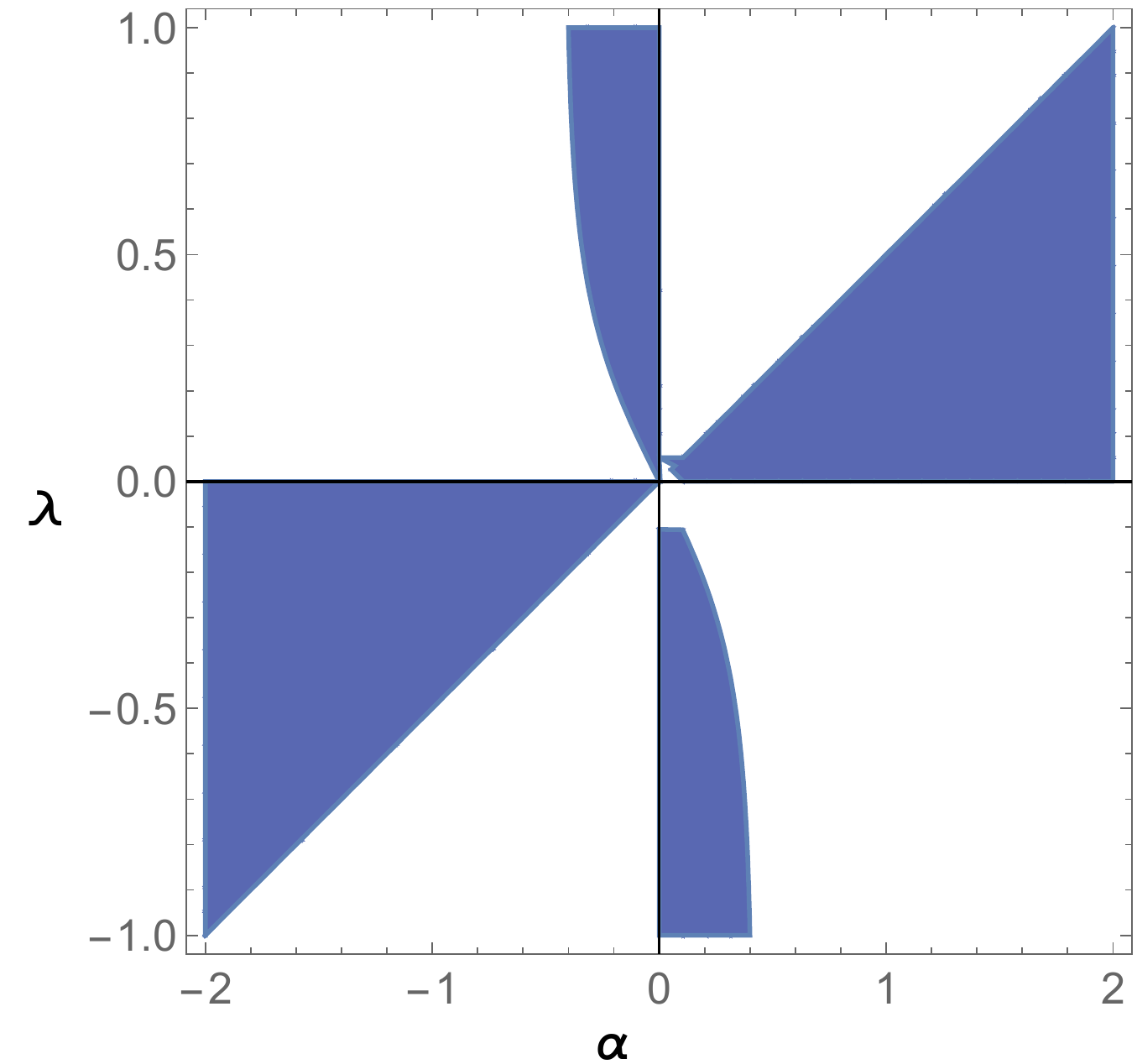}
\caption{The figure describes a possible interval for the $P_2$ critical point where the cosmological solution is physically viable and stable ($\epsilon=+1, w_m=0$, quintessence case).}
\label{fig:figp2qui}
\end{figure*}

\begin{figure}[t]
\centering
\includegraphics[width=0.4\textwidth]{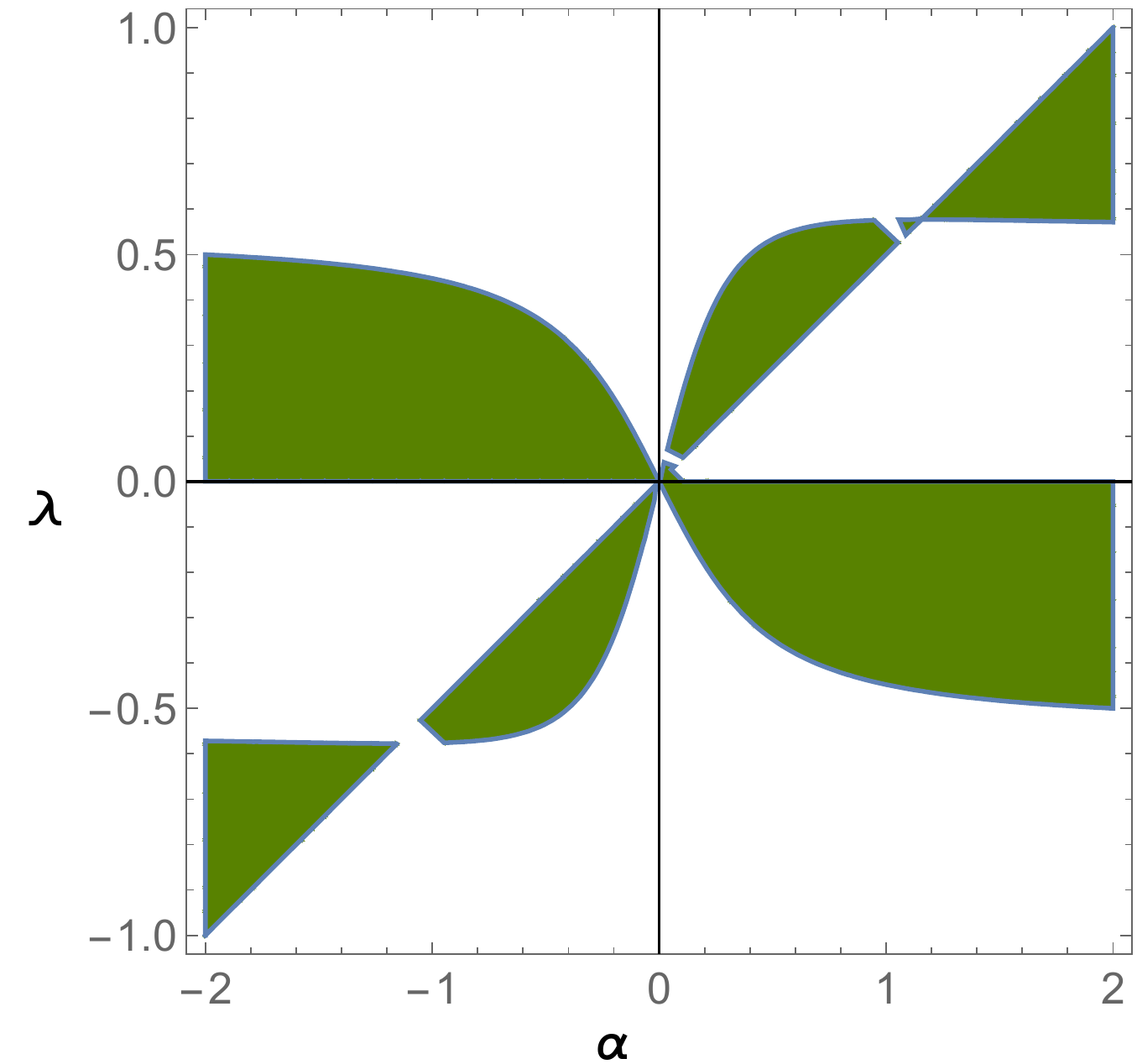}
\caption{The figure displays a region for the $P_2$ critical point where the cosmological solution is physically viable and stable ($\epsilon=-1, w_m=0$, phantom case).}
\label{fig:figp2pha}
\end{figure}

\begin{figure}
\centering
\includegraphics[width=0.4\textwidth]{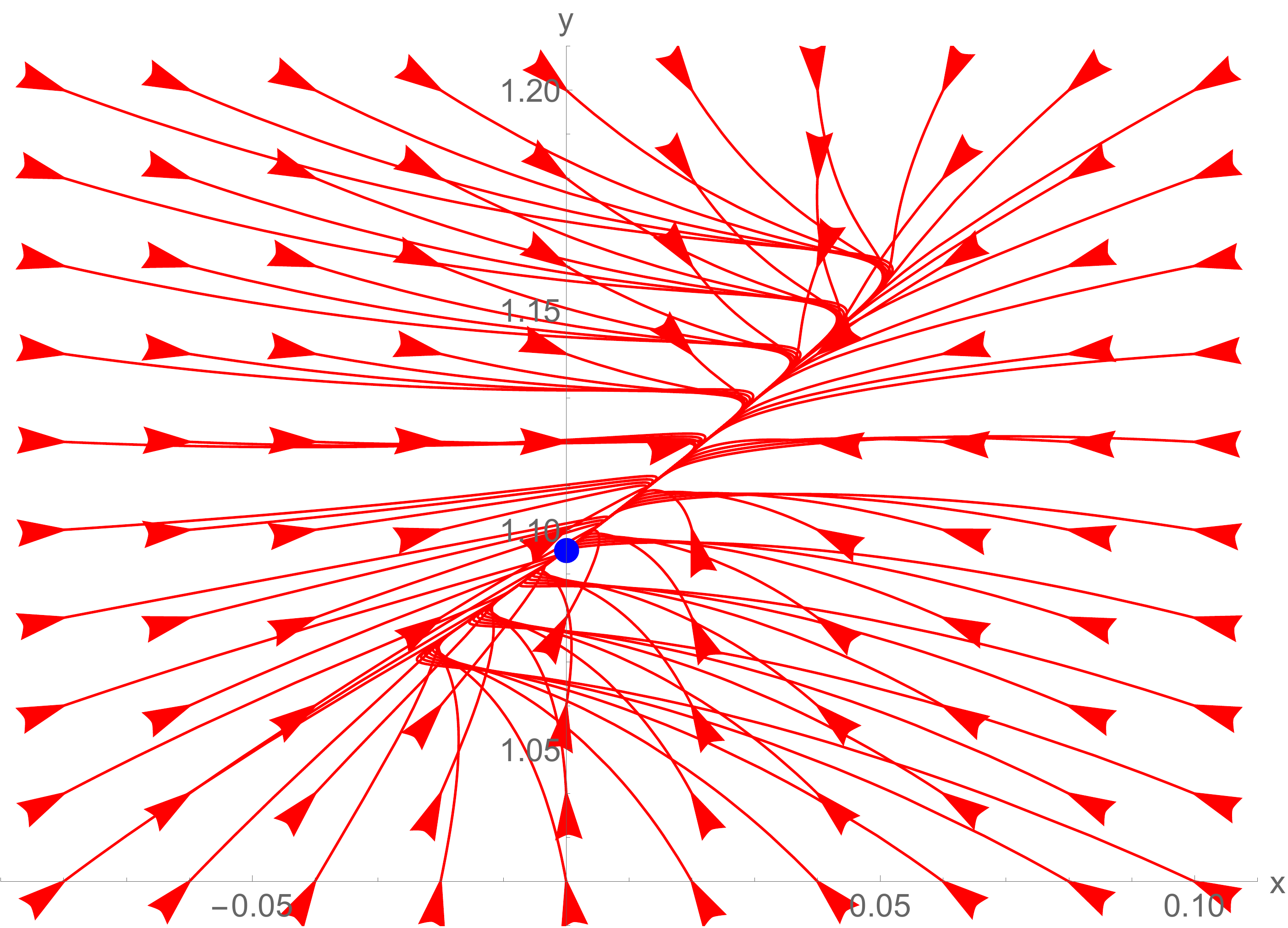}
\caption{The figure shows a numerical evolution towards the $P_2$ critical point. ($\epsilon=+1$, quintessence case, $\alpha=-1.5$, $\lambda=-0.5, w_m=0$).}
\label{fig:figp2ev}
\end{figure}

\begin{figure}
\centering
\includegraphics[width=0.4\textwidth]{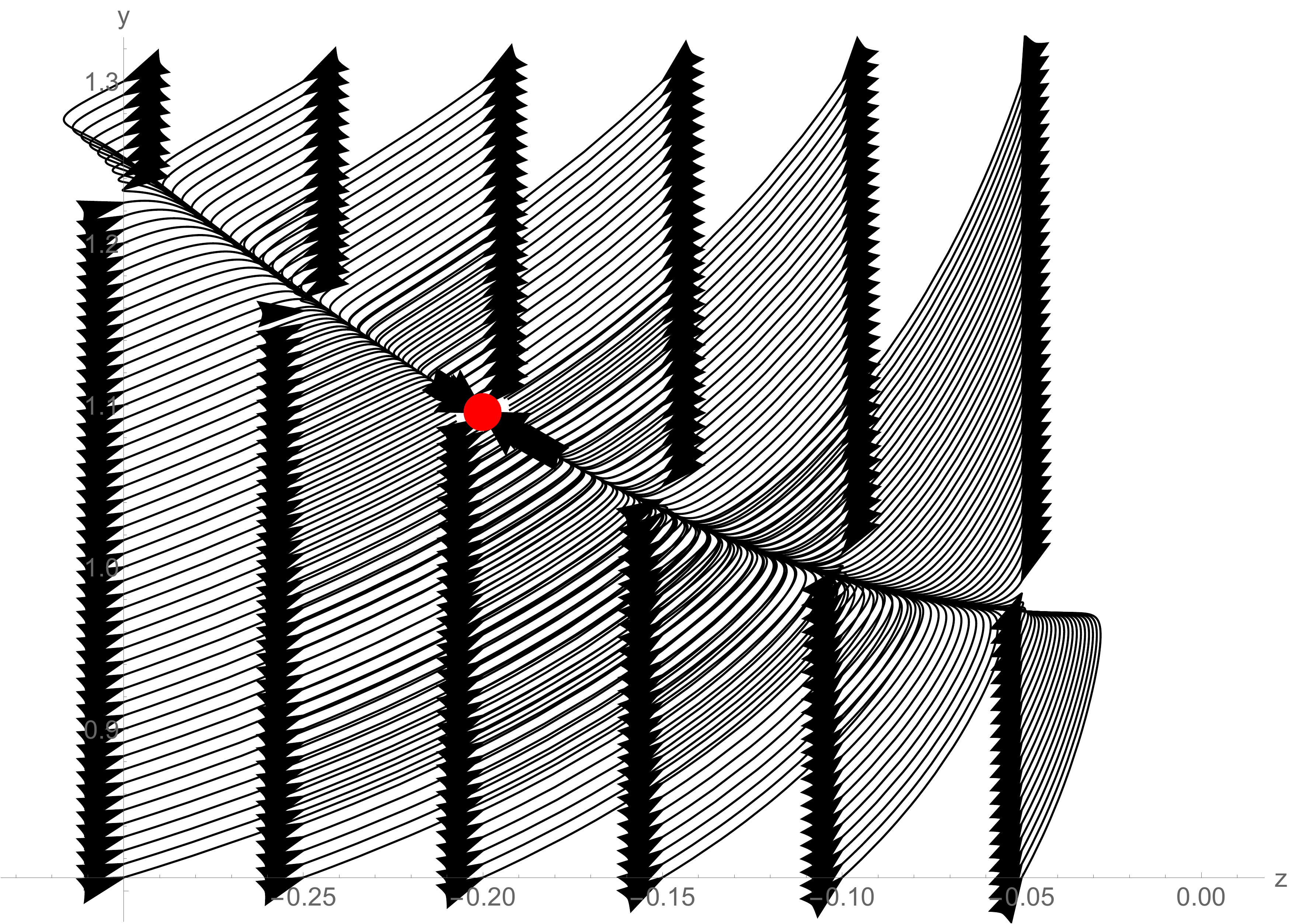}
\caption{The numerical evolution towards the $P_2$ solution. ($\epsilon=+1$, quintessence case, $\alpha=-1.5, w_m=0$, $\lambda=-0.5$).}
\label{fig:figp2e2v}
\end{figure}

 
The second critical point $P_2$ represents a cosmological solution characterized by physical effects from the potential energy and the non--minimal coupling component. The location in the phase space structure is influenced by the $\alpha $ and $\lambda$ parameters, embedding viable effects from the strength of the cubic coupling and the potential energy. For this solution the scalar field mimics a de Sitter regime, behaving as a cosmological constant. From a dynamical perspective the explicit form of the resulting eigenvalues are too cumbersome to be written in the manuscript, the analysis continuing with the presentation of some figures in some specific cases. At this point we also have to take into account the existence conditions related to a non zero denominator and the fact that the potential energy has to be positive, requiring $y \in R_{\ge 0}$. For the $P_2$ cosmological solution we have presented in Figs.~\ref{fig:figp2qui}--\ref{fig:figp2pha} some possible intervals for the variation of the $\lambda$ and $\alpha$ parameters which correspond to a stable scenario in the case of quintessence and phantom instances. The numerical evolution near the $P_2$ cosmological solution is represented in Figs.~\ref{fig:figp2ev}--\ref{fig:figp2e2v} in the stable scenario for specific initial conditions and values of the corresponding parameters. 

\onecolumngrid

\begin{figure}
\centering
\includegraphics[width=0.4\textwidth]{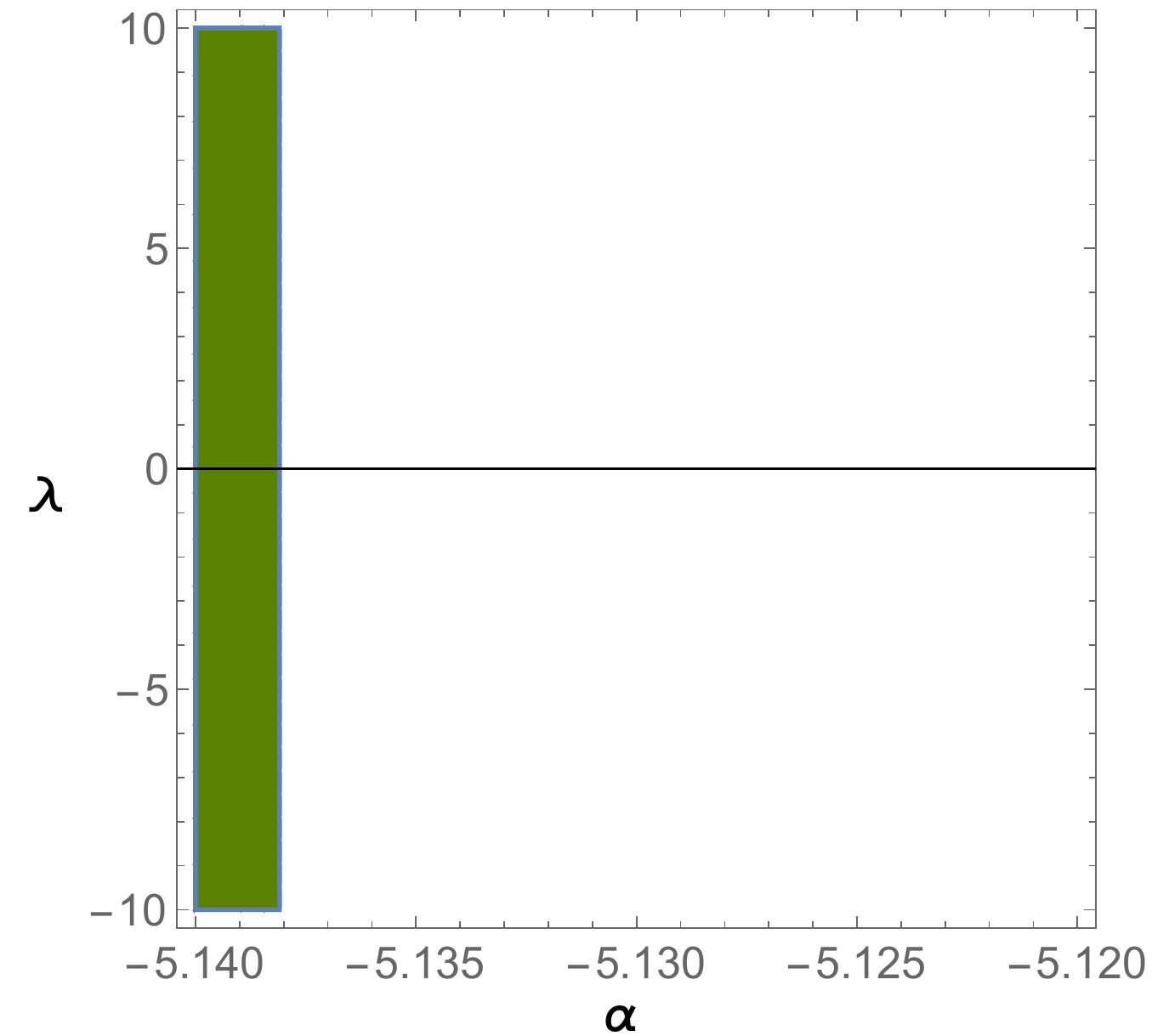}
\includegraphics[width=0.4\textwidth]{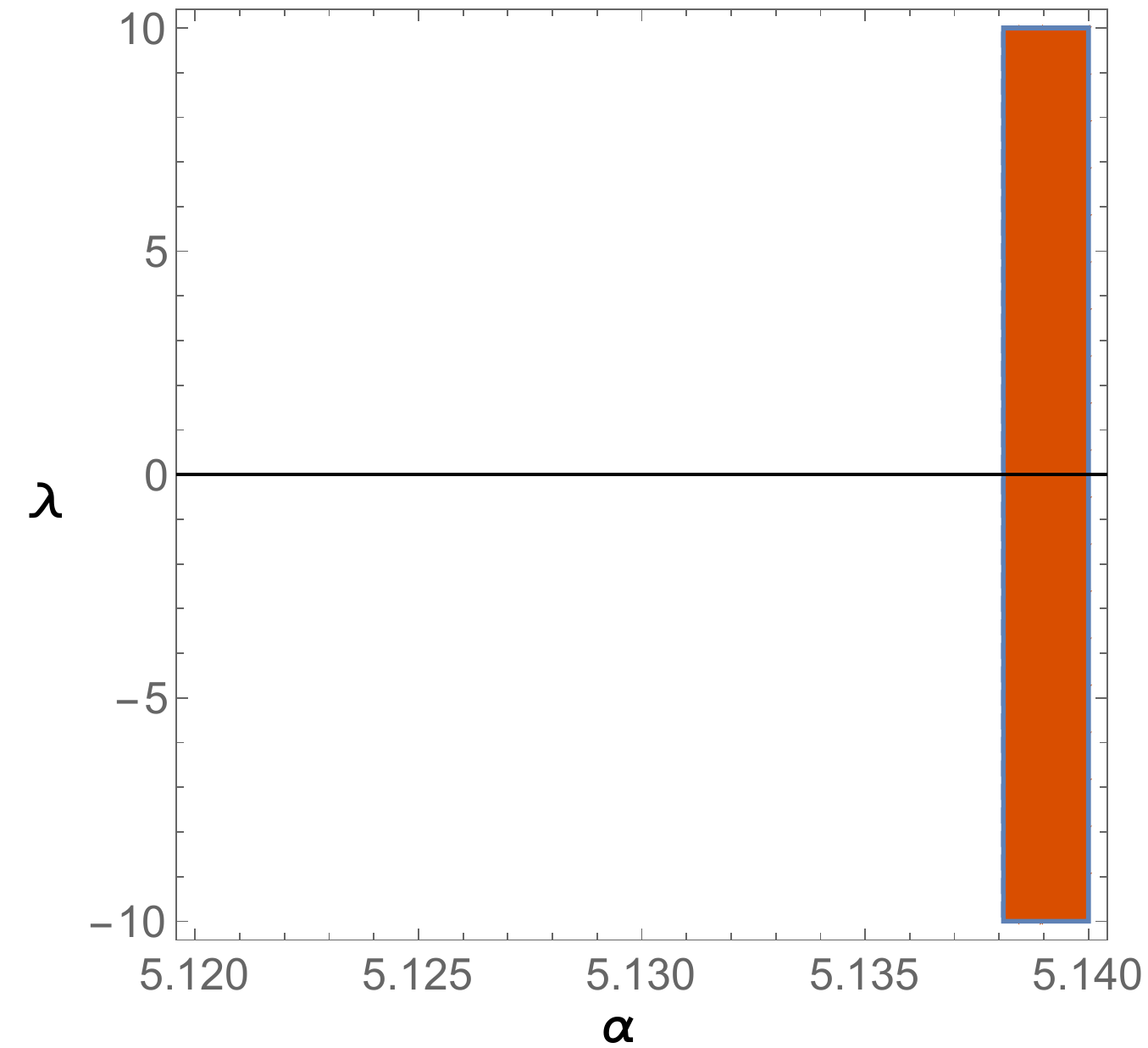}
\caption{The figure is showing specific intervals for the $P_3$ critical point where the cosmological solution is physically viable and saddle ($\epsilon=+1$, quintessence case).}
\label{fig:figp3a}
\end{figure}

\begin{figure}
\centering
\includegraphics[width=0.4\textwidth]{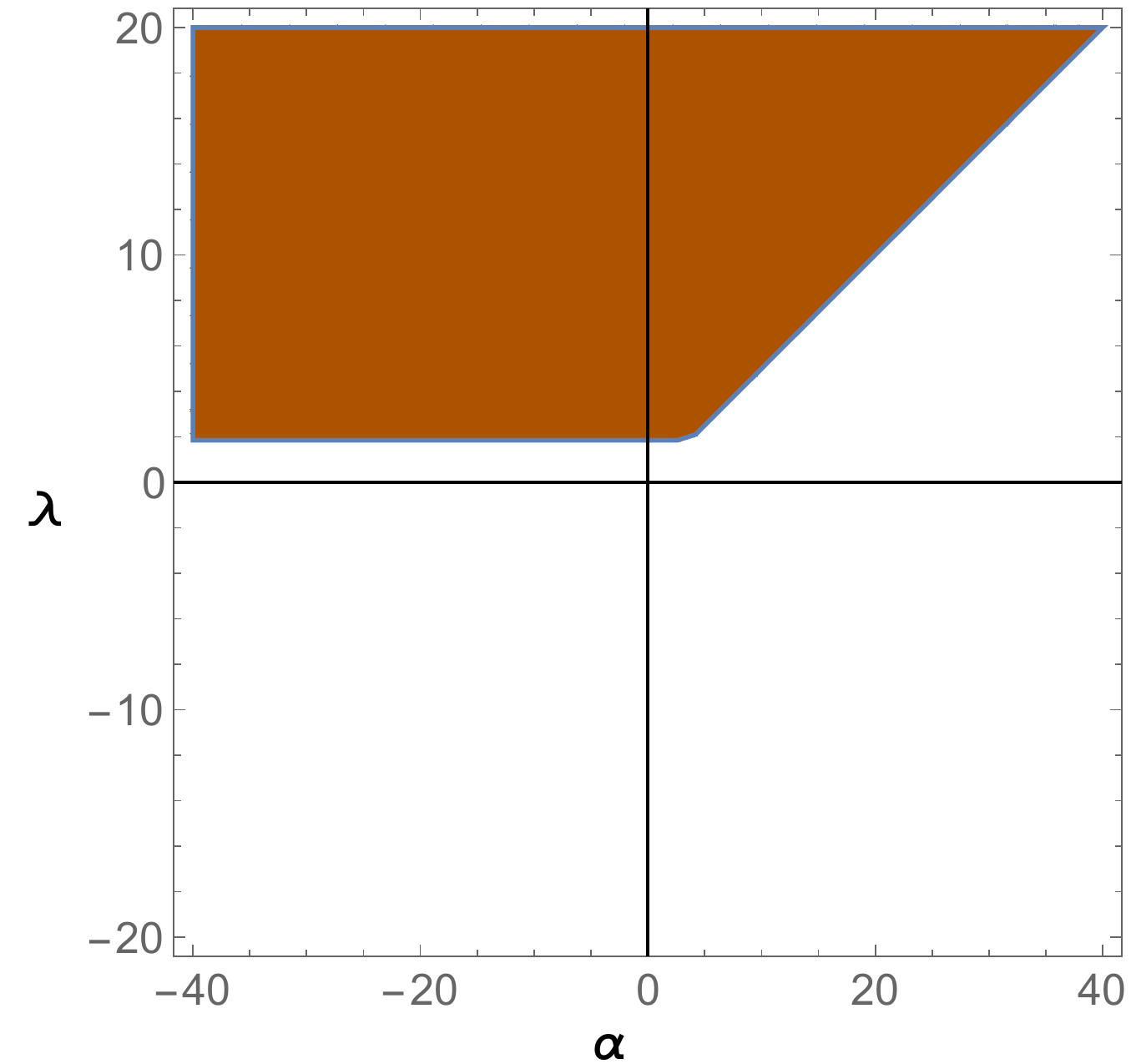}
\caption{The figure is showing specific intervals for the $P_7$ critical point where the cosmological solution is physically viable and stable spiral ($\epsilon=+1$, quintessence case, $w_m=0$).}
\label{fig:fig7q1}
\end{figure}

\begin{figure}
\centering
\includegraphics[width=0.4\textwidth]{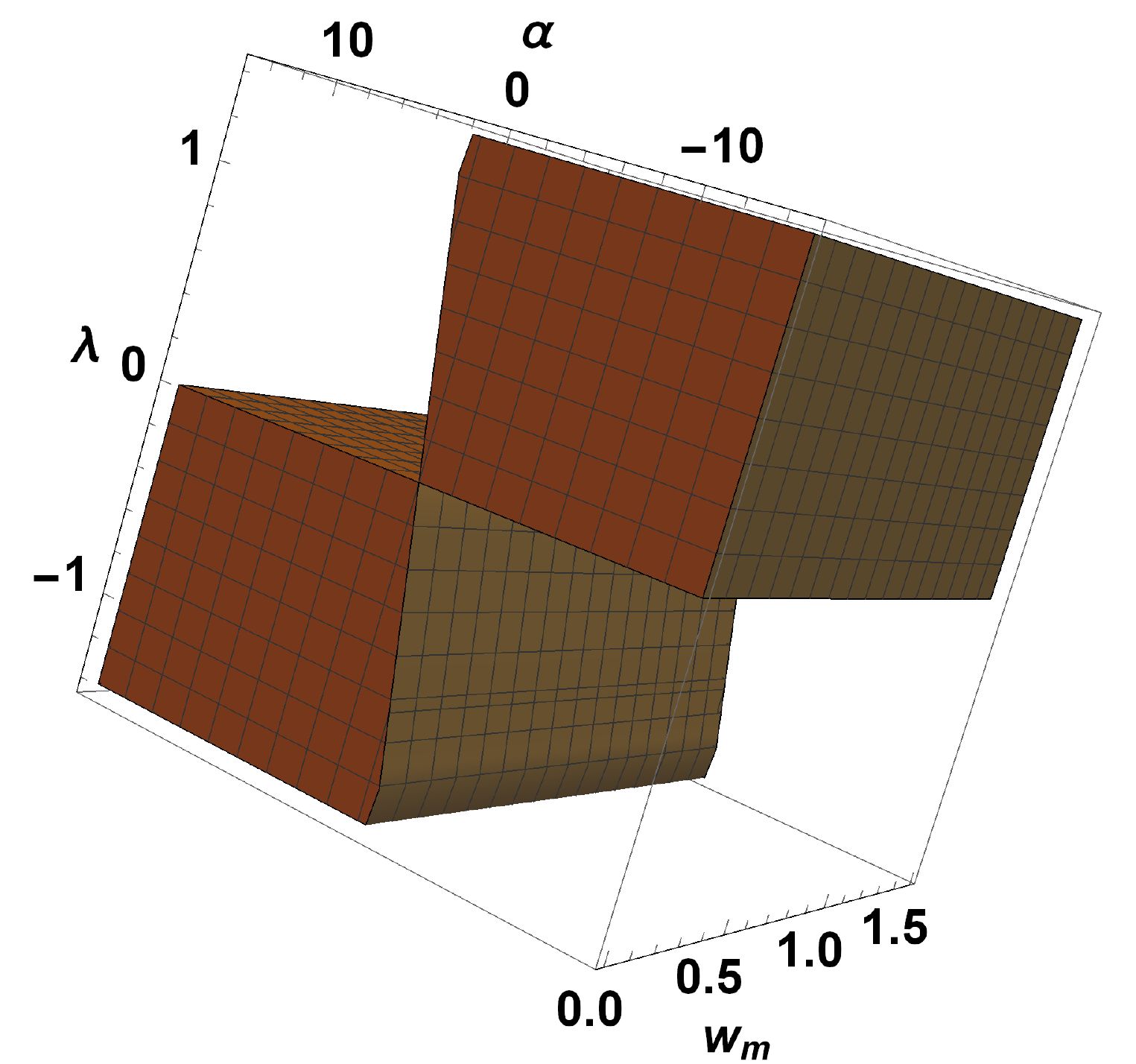}
\caption{The figure is showing specific intervals for the $P_8$ critical point where the cosmological solution is physically viable and stable, describing an accelerated epoch ($\epsilon=+1$, quintessence case).}
\label{fig:figp8}
\end{figure}

\twocolumngrid
\par 
The next critical point, the $P_3$ solution represents a scaling solution where the kinetic energy and the cubic coupling are affecting the physical features, the location in the phase space structure and the dynamical features. In the case where the matter component is a dust ($w_m=0$), the resulting eigenvalues have a simple expression as follows:
\onecolumngrid
\begin{multline}
    \Big[ 
    -\frac{75 \alpha ^3 \epsilon }{100 \alpha ^3 \epsilon -2016 \alpha  \epsilon ^2}+\frac{1512 \alpha  \epsilon ^2}{100 \alpha ^3 \epsilon -2016 \alpha  \epsilon ^2}-\frac{3 \sqrt{10625 \alpha ^6 \epsilon ^2-490800 \alpha ^4 \epsilon ^3+5576256 \alpha ^2 \epsilon ^4}}{100 \alpha ^3 \epsilon -2016 \alpha  \epsilon ^2},
    \\-\frac{75 \alpha ^3 \epsilon }{100 \alpha ^3 \epsilon -2016 \alpha  \epsilon ^2}+\frac{1512 \alpha  \epsilon ^2}{100 \alpha ^3 \epsilon -2016 \alpha  \epsilon ^2}+\frac{3 \sqrt{10625 \alpha ^6 \epsilon ^2-490800 \alpha ^4 \epsilon ^3+5576256 \alpha ^2 \epsilon ^4}}{100 \alpha ^3 \epsilon -2016 \alpha  \epsilon ^2},\frac{3}{2}-\frac{3 \lambda }{\alpha }
    \Big].
\end{multline}
\twocolumngrid
If we further consider the dust type for the matter component ($w_m=0$) and the quintessence case ($\epsilon=+1$), then the matter density parameter is equal to 
\begin{equation}
    \Omega_m=1-\frac{132}{5 \alpha ^2},
\end{equation}
respecting the viability conditions, i.e. $0 \le \Omega_m=1-\Omega_{\phi} \le 1$.
For the $P_3$ critical point we have represented in Fig.~\ref{fig:figp3a} possible intervals for the $\lambda$ and $\alpha$ parameters where the solution is physically viable, describing a saddle scenario from a dynamical perspective.

Further, if we take into account the dust type for the matter component ($w_m=0$) and the phantom case ($\epsilon=-1$), then the matter density parameter is equal to 
\begin{equation}
    \Omega_m=\frac{132}{5 \alpha ^2}+1,
\end{equation}
describing a solution which is not physically viable, not meeting the viability conditions presented above.

\par 
Next, the $P_4$ and $P_5$ critical points are kinetic solutions which are describing an era dominated by the kinetic energy of the scalar field, simulating a stiff--fluid scenario. This solution is physically viable only in the quintessence case, where $\epsilon=+1$. We should note that in the dust case ($w_m=0$) one of the eigenvalues is always positive. Hence, from a dynamical point of view this cosmological solution can be either saddle or unstable, depending on the sign and values of the $\lambda$ and $\alpha$ parameters, representing a particular epoch which has a limited interest in the analysis due to the non--negativity of the total effective equation of state.

\par 
The $P_6$ and $P_7$ critical points are characterized by the scalar field's kinetic and potential energy terms, simulating a matter dominated epoch as a background physical effect. The location in the phase space structure is affected by the sign of the kinetic energy, the strength of the potential, and the barotropic parameter $w_m$ which encodes effects from the pressure of the matter component. As can be noted, the $P_6$ solution is physically viable if we take into consideration that the $\lambda$ parameter is negative, while for the $P_7$ the $\lambda$ constant has to be positive. For the $P_6$ and $P_7$ cosmological solutions we have obtained the following eigenvalues:
\onecolumngrid
\begin{multline}
    \Big[\frac{3 (\alpha -2 \lambda ) \left(w_m+1\right)}{\lambda },\frac{3}{4} \left(w_m-1\right)-\frac{3 \sqrt{\lambda ^{10} \left(-\epsilon ^2\right) \left(w_m-1\right) \left(24 \epsilon  \left(w_m+1\right){}^2-\lambda ^2 \left(9 w_m+7\right)\right)}}{4 \lambda ^6 \epsilon },
    \\\frac{3 \sqrt{\lambda ^{10} \left(-\epsilon ^2\right) \left(w_m-1\right) \left(24 \epsilon  \left(w_m+1\right){}^2-\lambda ^2 \left(9 w_m+7\right)\right)}}{4 \lambda ^6 \epsilon }+\frac{3}{4} \left(w_m-1\right)\Big].
\end{multline}
\twocolumngrid
In what follows we shall present the physical and dynamical analysis for the $P_7$ solution which implies a positive sign for the $\lambda$ parameter. If we set the cold scenario ($w_m=0$), then the $y$ part of the solution depends on the square root of the $\epsilon$ parameters, viable from a physical point of view only in the quintessence case ($\epsilon=+1$). In this case the expressions for the eigenvalues acquire a more simpler form,
\onecolumngrid
\begin{equation}
    \Big[\frac{3 \alpha }{\lambda }-6,-\frac{3 \sqrt{24 \lambda ^{10}-7 \lambda ^{12}}}{4 \lambda ^6}-\frac{3}{4},\frac{3 \sqrt{24 \lambda ^{10}-7 \lambda ^{12}}}{4 \lambda ^6}-\frac{3}{4}\Big],
\end{equation}
\twocolumngrid
with the matter density parameter:
\begin{equation}
    s=1-\frac{3}{\lambda ^2}.
\end{equation}
A possible interval for the variation of the $\alpha$ and $\lambda$ where this solution is viable and stable spiral (with imaginary parts in the eigenvalues) is presented in Fig.~\ref{fig:fig7q1}. If we take into account a particular case where the $P_7$ critical point is purely stable (all the eigenvalues are real and negative), then we obtain the following conditions for the $\lambda$ and $\alpha$ parameters:
\onecolumngrid
\begin{equation}
    \left(\alpha \leq 2 \sqrt{3}\land \sqrt{3}<\lambda \leq 2 \sqrt{\frac{6}{7}}\right)\lor \left(2 \sqrt{3}<\alpha <4 \sqrt{\frac{6}{7}}\land \frac{\alpha }{2}<\lambda \leq 2 \sqrt{\frac{6}{7}}\right).
\end{equation}
\twocolumngrid
\par 
The last critical point, denoted as $P_8$ represents a scalar field dominated solution described by the kinetic and potential energies of the scalar field. The location in the phase space structure is affected mainly by the sign of the kinetic energy and the $\lambda$ parameter. In this case, for a canonical scalar field ($\epsilon=+1$) the effective equation of state can correspond to a quintessence regime. However, for negative $\epsilon$ the effective equation of state becomes super--accelerated, associated to a phantom regime. For this solution the expressions of the corresponding eigenvalues are simple as presented in Table~\ref{tab:table1}. We have displayed in Fig.~\ref{fig:figp8} a three dimensional region plot where the solution is viable and stable, describing an accelerated expansion era. Lastly, we note that the last three cosmological solutions $P_6, P_7, P_8$ are not affected by the specific cubic couplings and might appear \cite{Bahamonde:2017ize} in the phase space structure also in the minimal coupling case.

\section{Conclusions}
\label{sec:concluzii}

\par 
In this paper we have further extended the Einstein--Hilbert action by adding a scalar field non--minimally coupled in a generic manner to specific contractions of the Riemann tensor up to the cubic order. Taking into account specific interrelations between various constants for the cubic term, the resulting modified Friedmann relations associated to the gravitational sector become second order equations. The analysis presented here have taken into account the case of a positive and negative kinetic term, a scalar field model endowed with a potential energy of the exponential type. After deducing the Klein--Gordon equations and the modified Friedmann relations in the case of the Robertson--Walker metric, the analysis continue with the usage of linear stability theory in the case of an exponential coupling function associated to the term which contains the cubic contractions of the Riemann tensor. In this case we have introduced the phase space variables, approximating the dynamics for the cosmological model as an autonomous system of differential equations. Further, we have determined the associated critical points which can represent different stages in the evolution of the Universe, linearizing the equations for the dynamical system around the cosmological solutions.  
\par 
In the case of scalar fields (quintessence and phantom, respectively) non--minimally coupled to a component based on specific contractions of the Riemann tensor up to the cubic order we have observed that the phase space structure is rich in the dynamical features, containing eight critical points which are physically viable. The analysis showed the existence of five classes of critical points, offering possible trajectories associated to the evolution of the Universe as it is considered in the modern cosmological theories from an observational point of view. 
\par 
The first class of critical points represents a dynamical solution where the effective equation of state mimics a stiff--fluid scenario, an epoch which is not of great interest in the modern cosmology since it cannot explain the matter dominated epoch and the accelerated expansion near the de--Sitter regime at late times without fine--tuning the initial conditions in the complexity of the phase space structure. For this solution we note the domination of the dark energy component, en epoch characterized and driven mainly by the kinetic term, viable only for the canonical case. From a dynamical point of view this solution cannot be stable, it exists only in the quintessence scenario, denoting a saddle or unstable behavior.  
\par 
The second class of cosmological solutions is represented by different critical points which can explain the matter dominated epoch in the history of our Universe. In the analysis the origin of the phase space represents a matter dominated epoch characterized by the domination of the matter component, an era which is always saddle if we take into account that the pressure of the matter is absent. For this critical point we note that the scalar field is not driven the dynamics, having no manifestation from the kinetic and potential energy, or the coupling term. In our analysis we have observed that the structure of the phase space consists of three more cosmological solutions where the effective equation of state mimics a matter dominated epoch, critical points which are driven by various components of the scalar field non--minimally coupled to specific contractions of the Riemann tensor up to third order. For these critical points we note that the scalar field is driven the dynamical features through the kinetic and the potential energy terms, or the exponential coupling component, mimicking a matter dominated epoch which can be stable in certain scenarios. In these cases we have identified possible constraints for different specific parameters of our model from a dynamical point of view, leading to stable trajectories in the phase space.
\par 
A third class of cosmological solutions is represented by the de--Sitter epoch where the effective equation of state corresponds to the cosmological constant, explaining the late time evolution of our Universe near the phantom divide line boundary. This cosmological scenario is affected mainly by the potential energy and the exponential coupling function, without any influence from the kinetic energy term, an era where the scalar field can be regarded as frozen. In this case the dark energy component dominates in terms of density parameters and can lead to a stable or saddle comportment in various specific cases which have been analyzed in the attached figures. We note that the sign of the kinetic term associated to the scalar field is affecting mainly the form of the specific eigenvalues and the corresponding dynamics.
\par 
The last class of critical points is represented by a cosmological solution where we have obtained the accelerated expansion as a possible physical effect. The analysis showed the existence of one critical point which belongs to this class, a solution characterized by the kinetic and potential energy terms, without any influence from the coupling function in the location of the phase space structure. In the case of a negative kinetic energy the effective equation of state corresponds to a phantom origin, while for the canonical case it can correspond to a quintessence regime. For this solution we have identified proper regions associated to various parameters of our model which can lead to a stable regime capable of explaining the evolution towards the accelerated expansion as a main physical effect, confirming the viability of the present cosmological scenario.
\par 
The richness of the phase space structure in terms of physical features unfolded within this study are asserting the viability of this cosmological scenario which can explain the accelerated expansion and the existence of the matter dominated epoch, a feasible model from a dynamical point of view.
\par

\section{Acknowledgements}
 For the analytical study in this paper we have considered Wolfram Mathematica \cite{Mathematica} and the xAct package \cite{xact}.

\bibliography{article}
\bibliographystyle{apsrev}

\end{document}